\documentclass[fleqn,usenatbib,useAMS]{mnras}
\usepackage{newtxtext,newtxmath}

\usepackage{graphicx}	
\usepackage{amsmath}	
\usepackage{multicol}        
\usepackage{bm}		
\usepackage{pdflscape}	

\usepackage[normalem]{ulem} 
\usepackage[dvipsnames]{xcolor}
\usepackage{soul}

\usepackage{tikz}
\usetikzlibrary{calc, positioning}

\usepackage[percent]{overpic}

\usepackage{booktabs, multirow, makecell}

\setcellgapes{5pt}
\newcommand{\msol}{\, M_{\odot}}

\newcommand{\kms}{\,{\rm km \, s^{-1}}}
\newcommand{\pc}{\,{\rm pc}}
\newcommand{\kpc}{\,{\rm kpc}}

\newcommand{\oversim}[2]{\protect{\mbox{\lower0.5ex\vbox{%
   \baselineskip=0pt\lineskip=0.2ex
   \ialign{$\mathsurround=0pt #1\hfil##\hfil$\crcr#2\crcr\sim\crcr}}}}} 
\newcommand{\bb}[1]{\ifmmode \mbox{\boldmath $ #1$} \else  \mbox{\boldmath $#1$} \fi}
\def\3{{\ss} }

\def\c12{{1\over 2}}

\def\d{{\rm d}}   
   
\def\plusplus{\raise 0.3ex\hbox{${\scriptstyle ++}$}{}}

\def\and{{{\rm M}31}}

\begin{document}

\title[Friction vs. heating]{Dynamical friction vs. subhalo heating in Cold Dark Matter haloes}
\author[Pe\~narrubia et al.]{Jorge Pe\~narrubia$^{1,2,3}$\thanks{Email: jorpega@valer-f.es}, Pierfrancesco Di Cintio$^{4,5,6}$, Eduardo Vitral$^{3}$, Matthew G. Walker$^{7}$\\
$^1$Institute of Corpuscular Physics (IFIC), CSIC–Universitat de València, 46980 Paterna, Valencia, Spain\\
$^2$VALER, Calle Mayor, 83, 1, 12001 Castellón de la Plana, Spain\\
$^3$Institute for Astronomy, University of Edinburgh, Royal Observatory, Blackford Hill, Edinburgh EH9 3HJ, UK\\
$^4$Istituto dei Sistemi Complessi, Consiglio Nazionale delle Ricerche, Via Madonna del Piano 17, Sesto Fiorentino, 50019, Italy.\\
$^5$Osservatorio Astrofisico di Arcetri, Istituto Nazionale di Astrofisica, Largo Enrico Fermi 5, Firenze, 50025, Italy.\\
$^6$Istituto nazionale di Fisica Nucleare, Via G. Sansone 1, Sesto Fiorentino, 50019, Italy.\\
$^7$McWilliams centre for Cosmology and Astrophysics, Department of Physics, Carnegie Mellon University, Pittsburgh, PA 15213, USA
}
\maketitle  

\begin{abstract}
The orbital evolution of massive stellar systems embedded in dark matter
haloes is governed by competing processes. Dynamical friction removes orbital
energy, whereas fluctuations in the gravitational field generated by dark
matter subhaloes inject energy through stochastic heating. We investigate the
balance between these two mechanisms using analytical arguments and numerical
experiments. We show that, for a given host halo and subhalo population, there
exists a critical mass $M_{\rm crit}$ at which heating and friction balance. Clusters with masses $M_{\rm cl}>M_{\rm crit}$ lose orbital energy and sink towards
the centre of the halo, whereas those with $M_{\rm cl}<M_{\rm crit}$ gain
energy and expand outwards. At $M_{\rm cl}\simeq M_{\rm crit}$, the two
processes balance, leading to a cluster population that, on average, neither
sinks nor expands within the host halo. In CDM haloes, the critical mass is primarily
controlled by the upper end of the subhalo mass function. Since massive
subhaloes are intrinsically rare, the balance between heating and friction
shows substantial halo-to-halo scatter. For dark matter haloes in the mass
range associated with dwarf spheroidal galaxies (dSphs), we find
$M_{\rm crit}\gtrsim10^5\msol$, comparable to the masses of globular
clusters in the Fornax dSph. Stochastic heating by dark substructure may therefore significantly delay the orbital decay of globular clusters and help alleviate the Fornax timing problem.
\end{abstract}
\begin{keywords}
Galaxy: kinematics and dynamics; galaxies: evolution; Cosmology: dark matter.
\end{keywords}

\section{Introduction}\label{sec:intro}
By the beginning of the 20th century, the atomic theory of matter was remarkably successful, yet the physical reality of atoms and molecules remained a matter of heated debate. While some regarded them as genuine constituents of matter, others considered them convenient theoretical constructs whose existence could not be directly tested. Einstein's (1905) seminal work on Brownian motion showed how the properties of invisible molecules could be inferred from the stochastic motion of visible particles.

The current cosmological paradigm, Cold Dark Matter (CDM), similarly predicts the presence of large numbers of self-gravitating substructures within galaxies, known as {\it subhaloes}, that do not emit or absorb radiation and therefore remain invisible, i.e. `dark' (e.g. Benitez-Llambay \& Frenk 2020). Their presence may nevertheless be inferred through their gravitational perturbations on visible bodies, such as cold tidal streams (Ibata et al. 2002; Erkal \& Belokurov 2015; Bovy et al. 2017) and wide binary stars (Peñarrubia et al. 2010), through perturbations of strongly lensed images (Dalal \& Kochanek 2002; Vegetti \& Koopmans 2009), or indirectly through measurements of the small-scale matter power spectrum from the Lyman-$\alpha$ forest (Viel et al. 2013).

Dwarf spheroidal galaxies (dSphs) provide ideal laboratories to search for dark subhaloes. These galaxies host predominantly old stellar populations, contain little or no gas and---crucially---are embedded within extended DM haloes that dominate their gravitational potential (e.g. Simon 2019). In the absence of significant baryonic components, perturbations to stellar orbits therefore provide a clean probe of fluctuations in the underlying dark matter distribution.

Recently, Pe\~narrubia et al. (2025; hereafter P25) showed that dark subhaloes generate a fluctuating gravitational field that injects energy into stellar orbits. In dark-matter-dominated systems such as dSphs, subhalo heating drives a gradual expansion of the stellar distribution within the dark matter potential. 
This expansion leads to subtle fluctuations in the surface density of dSphs that may be detectable with forthcoming photometric surveys (Vitral, Pe\~narrubia \& Walker 2026).

In P25, stars are treated as massless tracers of the underlying DM potential. However, massive systems travelling through a collisionless DM background also experience dynamical friction (Chandrasekhar 1943), which removes orbital energy and drives them towards the central regions of the host. Bekenstein \& Maoz (1992; see also Pasquato \& Di Cintio 2020; Di Cintio \& Marcos 2025) showed that systematic drift and stochastic diffusion can be studied within a common framework, in which the same discrete population of massive particles gives rise to both effects. 

A long-standing example of this tension is provided by the globular cluster system of the Fornax dSph. Fornax hosts five\footnote{The nature of F6 is still debated (Wang et al. 2019; Pe\~narrubia et al. 2024) as it appears to be DM dominated (Pace et al. 2021). Similarly, the recently discovered F7 has an uncertain nature (Saifollahi et al. 2026).} globular clusters with masses of order $10^5\msol$ at projected distances of several hundred parsecs to $\sim 1.5\kpc$ from the galaxy centre, despite classical dynamical-friction estimates predicting orbital decay on timescales shorter than their ages (Tremaine 1976; Hernandez \& Gilmore 1998).
This so-called Fornax `timing problem' has often been interpreted as evidence for a low-density dark-matter core, in which dynamical friction may be strongly suppressed or stall altogether (Goerdt et al. 2006; Read et al. 2006). In general, the survival of the Fornax clusters may depend sensitively
on their initial orbital distribution (Boldrini, Mohayaee \& Silk 2019; Meadows et al. 2020),
while Boldrini, Mohayaee \& Silk (2020) showed that recently accreted clusters
embedded in their own DM minihaloes can induce a cusp-to-core transformation
and avoid orbital decay. Interestingly, numerical experiments have also shown that massive objects inside sufficiently low-density cores may gain orbital energy and migrate outwards through a process known as `dynamical buoyancy' (Cole et al. 2012). Recent theoretical work by Banik \& van den Bosch (2021, 2022) associates this behaviour with a reversal of the dynamical-friction torque within the core, such that buoyancy can counteract friction and lead to core stalling.
More generally, the orbital evolution of massive stellar systems through dynamical friction has been studied in a variety of contexts, including globular-cluster inspiral and multi-component collisionless systems (e.g. Bertin et al. 2003; Gnedin, Ostriker \& Tremaine 2014; Alessandrini et al. 2014).
These studies, however, largely focus on the systematic orbital evolution driven by dynamical friction. Below we show that if the dark halo also contains the population of subhaloes expected in CDM, stochastic heating provides an additional source of orbital energy that may counteract friction and modify the expected decay of the clusters.


\section{Energy balance equation}\label{Sec:energy}
Consider point mass $M_{\rm cl}$ moving within a CDM halo that contains $N$ subhaloes with inidividual masses $M$.
The equations that govern the motion of these objects can be written as (e.g. Bekenstein \&
Maoz 1992; Maoz 1993)
\begin{align}\label{eq:eqmot}
  \frac{\d^2\mathbf{r}}{\d t^2}=-\nabla\Phi_h+\mathbf{F}_{f,h}+\sum_{i=1}^N \mathbf{f}_i,
 \end{align}
 where $\Phi_h$ is the smooth halo potential, $\mathbf{F}_{f,h}$ is the dynamical-friction acceleration induced by the smooth halo, and $\mathbf{f}_i$ denotes the specific force induced by the $i$-th subhalo on the cluster. The combined force $\sum_{i=1}^N\mathbf{f}_i$ fluctuates randomly and acts as a noise term in the equations of motion.

Within a short time-interval $t$, a test particle $M_{\rm cl}$ experiences a variation of energy 
\begin{align}\label{eq:delEv}
 \langle \Delta E\rangle=\frac{\left\langle(\mathbf{v}+\Delta \mathbf{v})^2\right\rangle}{2}-\frac{\left\langle\mathbf{v}^2\right\rangle}{2}=
 \langle \mathbf{v}\cdot \mathbf{\Delta v}\rangle +\frac{1}{2} \langle |\Delta{\mathbfit v}|^2\rangle,
\end{align}
where $E=v^2/2+\Phi_h$ is the specific energy.
Therefore, the mean rate of change of orbital energy can be written as

\begin{align}\label{eq:dedt}
\left\langle\frac{{\rm d}E}{{\rm d}t}\right\rangle=
\left\langle\frac{{\rm d}E}{{\rm d}t}\right\rangle_f
+
\left\langle\frac{{\rm d}E}{{\rm d}t}\right\rangle_{d,\rm sub},
\end{align}
with brackets denoting averages over the cluster population.
The first term is negative and describes the systematic loss of orbital energy through dynamical friction, whereas the second is positive and accounts for the energy injected by stochastic fluctuations of the gravitational field generated by dark subhaloes. Below, we derive both terms separately and determine the conditions under which they balance.

\subsection{Cooling}

A cluster of mass $M_{\rm cl}$ moving with velocity $v=|\mathbf{v}|$ through an isotropic population of objects with mass $M$, number density $n$, and a Maxwellian velocity distribution with one-dimensional dispersion $\sigma$ feels a drag acceleration (Chandrasekhar 1943)

\begin{align}\label{eq:df}
\frac{{\rm d}\mathbf{v}}{{\rm d}t}
=-4\pi G^2 n\,M\,(M_{\rm cl}+M)\ln(\Lambda_f)
\frac{\hat{\mathbf v}}{v^2}
\left[
{\rm erf}(X)
-\frac{2X}{\sqrt{\pi}}\exp(-X^2)
\right],
\end{align}
where $X=v/(\sqrt{2}\sigma)$ and $\ln(\Lambda_f)$ is the corresponding Coulomb logarithm. In the limit $M\ll M_{\rm cl}$ Equation~(\ref{eq:df}) recovers Chandrasekhar (1943)'s dynamical friction.

Consider now an ensemble of clusters that follow a Maxwellian velocity distribution with one-dimensional dispersion $\sigma_{\rm cl}$. Averaging~(\ref{eq:df}) over the cluster velocities gives

\begin{align}\label{eq:dele_f}
\left\langle\frac{{\rm d}E}{{\rm d}t}\right\rangle_f
=
-4\sqrt{2\pi}G^2\,n\,M(M_{\rm cl}+M)\ln(\Lambda_f)
\frac{\sigma_{\rm cl}^2}
{(\sigma_{\rm cl}^2+\sigma^2)^{3/2}}.
\end{align}

We can now split the contribution to~(\ref{eq:dele_f}) from smooth halo particles and subhaloes.
For the microscopic particles that constitute the smooth halo,
$M=m\ll M_{\rm cl}$ and $n(r)m=\rho_h=M_h\,g(r)$, where $M_h$ is the total halo mass and $\int\d^3r\,g(r)=1$. Equation~(\ref{eq:dele_f}) reduces to
\begin{align}\label{eq:dele_f_smooth}
\left\langle\frac{{\rm d}E}{{\rm d}t}\right\rangle_{f,\rm h}
=
-4\sqrt{2\pi}G^2M_{\rm cl}M_h\,g
\ln(\Lambda_{f,\rm h})
\frac{\sigma_{\rm cl}^2}
{(\sigma_{\rm cl}^2+\sigma^2)^{3/2}}.
\end{align}
For subhaloes, the factor $M(M_{\rm cl}+M)$ reveals two distinct regimes. For $M\ll M_{\rm cl}$ the drift scales as $MM_{\rm cl}$, as in conventional dynamical friction, whereas for $M\gg M_{\rm cl}$ it scales as $M^2$ and becomes independent of the cluster mass. The latter represents the negative first-moment drift associated with discrete subhalo scattering. It has the same mass dependence as the velocity diffusion discussed below, but enters the energy balance with a negative sign. The corresponding cooling rate for a subhalo mass function is calculated below together with the stochastic-heating term.

\subsection{Heating}\label{sec:heating}
For a population of Hernquist (1990) spheres with mass $M$ and scale radius $c$ the variance of the velocity impulses is (Pe\~narrubia 2019; hereafter P19) 
\begin{align}\label{eq:delv2_long}
  \langle |\Delta{\mathbfit v}|^2\rangle =t\,\sqrt{\frac{32\pi^3}{3\langle v_{\rm rel}^2\rangle}}(GM)^2\,n\,\big[\ln (D/c) -1.9\big].
\end{align}
where $D=(2\pi \,n)^{-1/3}$ is the average distance between subhaloes and $n$ their number density. 
Averaging over the Maxwellian subhalo and cluster velocities gives
$\langle v_{\rm rel}^2\rangle=3(\sigma_{\rm cl}^2+\sigma^2)$. 
 Equation~(\ref{eq:delv2_long}) exhibits a classical stochastic behaviour, where the variance increases in proportion to the length of the time-interval, $\langle|\Delta{\mathbfit v}|^2 \rangle\propto t$.
As originally pointed out by Chandrasekhar (1941a,b), the `Coulomb logarithm' $\ln (\Lambda_{d,\rm sub})\equiv \ln(D/c)$, diverges in the point-mass limit $c/D\to 0$. Using numerical experiments, P19 found that the median value of the Coulomb logarithm flattens at a maximum value $\langle \ln(\Lambda_{d,
\rm sub})\rangle \approx 8.2$ for $c/D\lesssim 10^{-3}$ (see their Fig. 3). 
In what follows, we assume that $\rm ln (\Lambda_{d,\rm sub})$ is constant.
The heating rate term appearing in~(\ref{eq:delEv}) becomes

\begin{align}\label{eq:dele_h}
\left\langle\frac{{\rm d}E}{{\rm d}t}\right\rangle_{d,\rm sub}
=\frac{2\sqrt{2}\pi^{3/2}}{3}(GM)^2\,n\,\big[\ln (\Lambda_{d,\rm sub}) -1.9\big]\frac{1}{(\sigma_{\rm cl}^2+\sigma^2)^{1/2}}.
\end{align}

\subsection{Subhalo mass function}\label{sec:massfun}
Following Ciotti (2010), the frictional contribution from a spectrum of field-particle masses is obtained by integrating the drag over the mass function.
Following Han et al. (2016), one can express the number density of subhaloes in the mass range $M,M+\d M$ within a volume element $r,r+\d^3r$ as
\begin{align}\label{eq:dndm}
\frac{\d n}{\d M}=\frac{\d^4 N}{\d M\d^3 r}=\frac{\d N}{\d M}g(r),
\end{align}
where $g(r)$ follows the same profile as the smooth DM component and 
\begin{align}\label{eq:dNdM}
\frac{\d N}{\d M}=B_0\bigg(\frac{M}{M_\odot}\bigg)^{-\alpha},
\end{align}
is a power-law mass function with index $\alpha=1.9$ (Springel et al. 2008). The mass subhalo function is set by the normalization $B_0$, which has units of $M_\odot^{-1}$, and the mass range $(M_1,M_2)$. Physically, the low-mass end of the mass function ($M_1$) is mainly set by the CDM particle mass (e.g. Green et al. 2005), and it is safe to assume that $M_1\lll M_2$.

Integrating over the subhalo mass function~(\ref{eq:dndm}) and~(\ref{eq:dNdM}), and taking the time derivative yields
\begin{align}\label{eq:dele_h_sub}
\left\langle\frac{{\rm d}E}{{\rm d}t}\right\rangle_{d,\rm sub}=\frac{2\sqrt{2}\pi^{3/2}}{3} G^2 g\,B_0\,M_\odot^\alpha [\ln(\Lambda_{d,\rm sub})-1.9]\frac{1}{(\sigma_{\rm cl}^2+\sigma^2)^{1/2}} \\\nonumber
\times
\frac{ 1}{3-\alpha}[M_2^{3-\alpha}-M_1^{3-\alpha}].
\end{align}
Similarly, the cooling rate~(\ref{eq:dele_f}) averaged over the subhalo mass function yields
\begin{align}\label{eq:dele_f_sub}
\left\langle\frac{{\rm d}E}{{\rm d}t}\right\rangle_{f,{\rm sub}}
=&
-4\sqrt{2\pi}G^2 \,g\,B_0\msol^\alpha
\ln(\Lambda_{f,\rm sub})
\frac{\sigma_{\rm cl}^2}
{(\sigma_{\rm cl}^2+\sigma^2)^{3/2}}
\nonumber\\
&\times
\left[
M_{\rm cl}
\frac{M_2^{2-\alpha}-M_1^{2-\alpha}}
{2-\alpha}
+
\frac{M_2^{3-\alpha}-M_1^{3-\alpha}}
{3-\alpha}
\right].
\end{align}
As expected, the heating and cooling rates are directly proportional to the normalization of the subhalo mass function, $B_0$, but have different velocity dependencies. 
To gain physical insight, in what follows we set $B_0$ using the total fraction of halo mass contained
in substructure
\begin{align}\label{eq:fsub}
f_{\rm sub}
&\equiv
\frac{1}{M_h}
\int_{M_1}^{M_2}
\d M\,M\,\frac{\d N}{\d M}.
\end{align}
For $\alpha\ne2$, Equation~(\ref{eq:fsub}) gives
\begin{align}\label{eq:B0}
B_0
&=
\frac{f_{\rm sub}}
{M_\odot^\alpha M_h^{1-\alpha}}
\frac{2-\alpha}
{\xi_2^{2-\alpha}-\xi_1^{2-\alpha}}, \qquad \xi\equiv{M}/{M_h}.
\end{align}

\subsection{Critical mass}\label{sec:mcrit}
We now compute the cluster mass that balances heating and cooling. 
From Equations~(\ref{eq:dele_f_smooth}),~(\ref{eq:dele_h_sub}) and~(\ref{eq:dele_f_sub}), it is straightforward to show that the condition $\langle\d E/\d t\rangle=\langle\d E/\d t\rangle_{f,\rm h}+\langle\d E/\d t\rangle_{f,\rm sub}+\langle\d E/\d t\rangle_{d,\rm sub}=0$ in Equation~(\ref{eq:dedt})
is satisfied for clusters with a mass
\begin{align}\label{eq:mcrit}
M_{\rm crit}
=
\left[
\frac{\pi}{6}C_d h_\sigma-C_{f}
\right]
\frac{f_{\rm sub}}{1+C_{f}f_{\rm sub}}
 f_\alpha\,M_h,
\end{align}
where $f_\alpha=(2-\alpha)/(3-\alpha)(\xi_2^{3-\alpha}-\xi_1^{3-\alpha})/(\xi_2^{2-\alpha}-\xi_1^{2-\alpha})$
,$C_d \equiv
(\ln(\Lambda_{d,\rm sub})-1.9)/\ln(\Lambda_{f,\rm h})$, and $C_f \equiv
\ln(\Lambda_{f,\rm sub})/\ln(\Lambda_{f,\rm h})$ are dimensionless quantities. 
The factor  $h_\sigma=1+\sigma^2/\sigma_{\rm cl}^2$ is a local quantity, which in general varies as a function of radius. 
Recall that clusters with $M_{\rm cl}<M_{\rm crit}$
are dominated by stochastic heating and tend to expand within the host
halo, whereas systems with $M_{\rm cl}>M_{\rm crit}$ are dominated by
dynamical friction and sink towards the centre. 

Equation~(\ref{eq:mcrit}) highlights a few interesting properties of the
heating--friction balance. 
First, this Equation assumes that subhaloes trace the density profile of the smooth halo, $\eta(r)=g_{\rm sub}(r)/g_h(r)=1$. For $\eta\ne 1$, the relevant quantity controlling the heating--friction balance in~(\ref{eq:mcrit}) is the local abundance of substructure relative to the smooth halo, $f_{\rm sub}^{\rm loc}(r)\equiv\eta(r)f_{\rm sub}$. Hence, if subhaloes are less centrally concentrated than the smooth dark matter, $M_{\rm crit}$ decreases towards regions where $\eta(r)<1$.
Second, a positive critical mass exists provided that $\ln\Lambda_{f,\rm sub}<(\pi/6)h_\sigma[\ln(\Lambda_{d,\rm sub})-1.9)]$. 
 Thus, the existence of $M_{\rm crit}$ is determined by the competition between stochastic heating and the $M^2$ cooling term generated by the subhalo population itself. For $\sigma_{\rm cl}\simeq\sigma$ ($h_\sigma\simeq2$) and $\ln(\Lambda_{d,\rm sub})=8.2$ (P19), a positive critical mass exists provided $\ln(\Lambda_{f,\rm sub})\lesssim6.6$, independently of $\ln(\Lambda_{f,\rm h})$.
 Third, $M_{\rm crit}$ diverges as $\sigma_{\rm cl}\to 0$ ($h_\sigma\to\infty$), because
 the average frictional energy-loss rate vanishes for arbitrarily cold tracer populations while stochastic heating remains finite. In general, the average value of $\sigma_{\rm cl}$ is set by the spatial distribution of tracers in the halo potential through the virial theorem (e.g. Errani et al. 2018; Splawska et al. 2026). For cluster populations deeply segregated within the halo, we expect $\sigma_{\rm cl}\lesssim\sigma$ ($h_\sigma\gtrsim2$).

The fact that the slope of the CDM mass function is $\alpha<2$ has important consequences. 
Adopting $\alpha=1.9$ and defining $q\equiv\xi_1/\xi_2$,
Equation~(\ref{eq:mcrit}) becomes
\begin{align}\label{eq:mcrit_CDM}
M_{\rm crit}^{\rm CDM}
=
\frac{1}{11}
\left[
\frac{\pi}{6}C_d h_\sigma-C_f
\right]
\frac{1-q^{1.1}}{1-q^{0.1}}
\frac{f_{\rm sub}}{1+C_f f_{\rm sub}}
\,M_2.
\end{align}
Thus, for a CDM-like mass function, the heating--friction balance is primarily controlled by the upper end of the subhalo mass function, $M_2$. 
However, because $\alpha=1.9$ lies close to the critical slope
$\alpha=2$, the slope where the total subhalo mass per
logarithmic mass interval is constant, $M_{\rm crit}^{\rm CDM}$ has a
non-negligible dependence on the range of the mass function
through $q=M_1/M_2=\xi_1/\xi_2$, even though the stochastic heating itself is
dominated by the most massive subhaloes. 
In contrast, in the equal-mass limit $M_1\to M_2=M$, $f_\alpha M_h\to M$, the critical mass is set by the individual subhalo mass rather than by the upper end of the mass function. At fixed $f_{\rm sub}$, a population of low-mass perturbers, e.g. MACHOs, can therefore produce significantly less heating than a CDM subhalo mass spectrum. This is consistent with the granularity-induced suppression of orbital decay found by Di Cintio \& Marcos (2025), whose stalling radius depends on the tracer-to-field-particle mass ratio.


\section{Numerical experiments}\label{sec:num}
\subsection{Set up}
We adopt the numerical set-up introduced in P25 to integrate the equations of motion~(\ref{eq:eqmot}), with one important modification: rather than massless stellar tracers, here we consider point-mass clusters with mass $M_{\rm cl}$ that are subject both to gravitational perturbations from
dark subhaloes and to Chandrasekhar's dynamical friction from the smooth halo. For the latter, we adopt a constant Coulomb
logarithm
$\ln(\Lambda_{f,\rm h})=2.1$ (Just \& Pe\~narrubia 2005).

Our main goal is to isolate the competition between stochastic heating and dynamical friction. To this end, subhalo populations are treated as a statistically stationary source of gravitational fluctuations: these objects orbit within the host potential, but their masses and orbital distribution do not evolve as a result of dynamical friction, interactions with other subhaloes or with individual clusters. Thus, clusters do not perturb the subhalo orbits. However, their motion relative to the subhalo population introduces an anisotropy in the distribution of encounters, which generates a systematic velocity drift as well as stochastic heating.
We also make further simplifications: the host halo is spherical and time-independent, clusters are represented as point masses, and external tides, mergers and baryonic processes are neglected.
The experiments therefore describe the response of clusters to a statistically stationary fluctuating background. Notice that in a cosmological setting, subhaloes are accreted at different times, and experience dynamical friction and tidal mass loss, causing the amplitude and spectrum of gravitational fluctuations to vary with time. We come back to these points in \S\ref{sec:discussion}.

\begin{figure}
\begin{center}
\includegraphics[width=86mm]{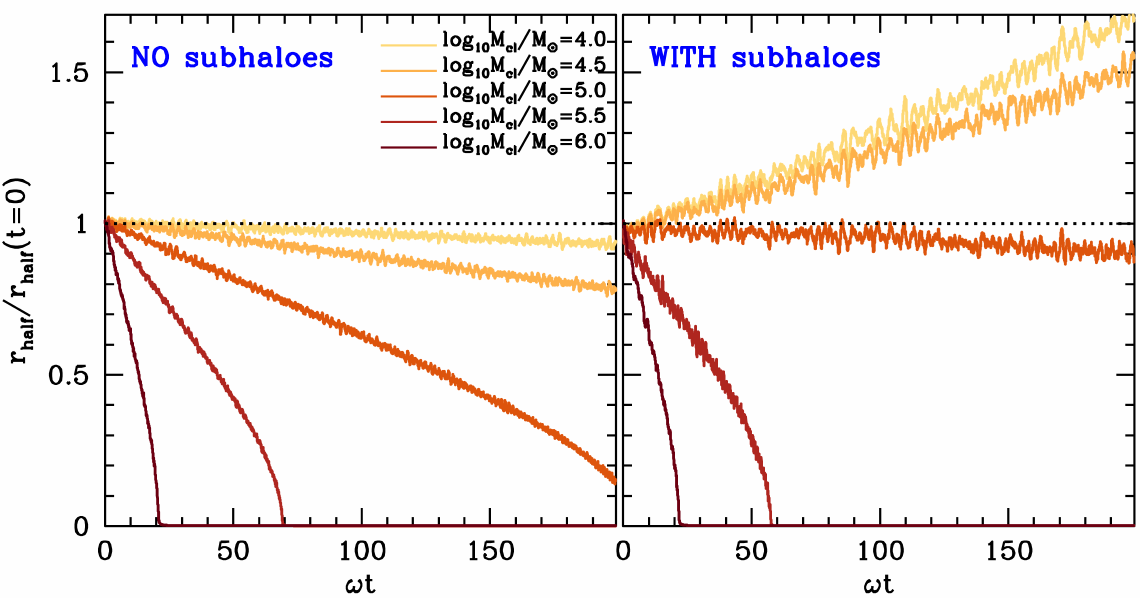}
\end{center}
\caption{Evolution of the half-light radius of $10^4$ point-mass clusters of mass $M_{\rm cl}$. Left: smooth Hernquist halo with $M_h=10^9\msol$ and $r_h=2.26\kpc$. Right: same halo containing $N=151$ subhaloes with $(\xi_1,\xi_2)=(10^{-4},0.03)$, where $\xi=M/M_h$ (see text). Stochastic heating causes the lightest clusters to expand.}
\label{fig:rhalf_2}
\end{figure}

The host galaxies are represented by Hernquist (1990) dark matter
potentials with masses
$M_h=\{3\times10^8,10^9,10^{10}\}\msol$ and scale radii chosen as in
P25 to reproduce the characteristic densities of Local Group dwarf
spheroidal galaxies (Strigari et al. 2007; Pe\~narrubia et al. 2008;
Kravtsov 2010; Errani et al. 2018). Unless stated otherwise, we adopt
$M_h=10^9\msol$ and $r_h=2.26\kpc$ as our fiducial model.

For each host halo, subhalo masses are drawn from the CDM mass
function~(\ref{eq:dNdM}) with $\alpha=1.9$ and a fixed normalization $B_0$ derived from the Aquarius simulations (Springel et al. 2008, see P25).The separability of~(\ref{eq:dndm}) implies that all subhalo mass bins share the same normalized spatial profile and, in equilibrium, the same velocity distribution.
We consider two mass ranges,
$(\xi_1,\xi_2)=(10^{-4},0.03)$ and $(10^{-5},0.03)$, containing
$N=151$ and $1211$ subhaloes, respectively. 

Subhaloes spatially trace the Hernquist profile of the host and are
represented by exponentially truncated NFW-like density profiles, with
structural parameters determined following Errani \& Navarro (2021) and
Aguirre-Santaella et al. (2023). Their initial velocities are drawn from
an Osipkov--Merritt distribution function (Osipkov 1979; Merritt 1985). We refer the reader to P25 for further details of
the subhalo initialization and numerical integration scheme.

We represent the cluster distribution by an ensemble of independent tracer orbits. Each cluster is evolved separately in the same statistical model of the halo and does not interact gravitationally with the other clusters. The ensemble should therefore be interpreted as a Monte Carlo sampling of the cluster distribution function.
The cluster population initially follows a Plummer (1911) profile with half-light radius $r_{\rm half}=790\pc$ chosen to match the spatial scale of the Fornax dSph (e.g. McConnachie 2012). The host DM halo masses and scale radii are chosen such that the average one-dimensional velocity dispersion of the cluster population is $\sigma_{\rm cl}\approx11\kms$ in all three halo models (see P25). The average one-dimensional velocity of the subhalo population is $\sigma=\sqrt{GM_h/(18\,r_h)}$ (Hernquist 1990). For the three halo masses considered here, $M_h/M_\odot=3\times10^8,10^9$, and $10^{10}$, this gives $\sigma=9.8,10.3$, and $15.4\kms$, respectively. Thus, in our models, clusters and subhaloes have a comparable velocity dispersion. Although both quantities generally vary with radius, we adopt $\sigma_{\rm cl}\sim\sigma$ for comparison with the numerical experiments below.

\begin{figure}
\begin{center}
\includegraphics[width=86mm]{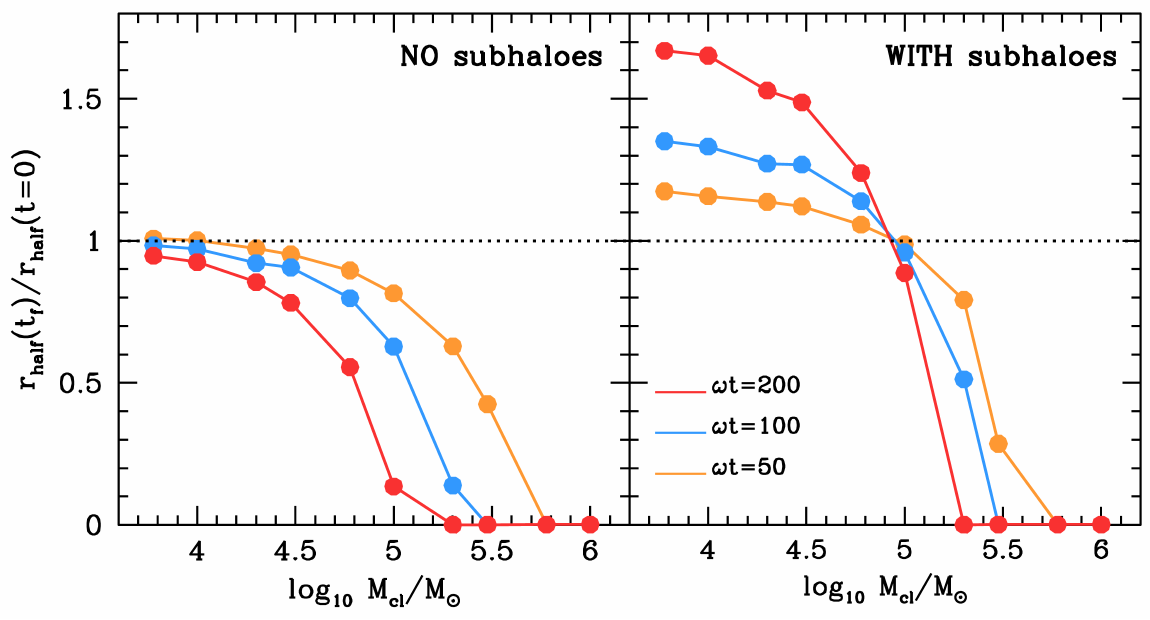}
\end{center}
\caption{Evolution of the half-light radius of the cluster populations plotted in Fig.~\ref{fig:rhalf_2} a function of  cluster mass. In halo models with subhaloes, the lines approximately cross at the critical mass, $M_{\rm cl}\approx M_{\rm crit}$.}
\label{fig:rhalf_tf}
\end{figure}

\subsection{Results}

Fig.~\ref{fig:rhalf_2} shows the evolution of the half-light radius of the
cluster population (i.e. the radius enclosing half of the clusters)  in the fiducial halo with $M_h=10^9\msol$. In the smooth model, dynamical
friction causes the population to contract at a rate that increases with
$M_{\rm cl}$. The addition of subhaloes changes this behaviour qualitatively:
clusters with $M_{\rm cl}\lesssim10^5\msol$ expand through stochastic
heating, whereas more massive clusters sink as dynamical friction dominates.
Close to $M_{\rm cl}\sim10^5\msol$, the competing effects approximately
balance and the half-light radius remains nearly constant.

Fig.~\ref{fig:rhalf_tf} illustrates this transition more directly by
showing the half-light radius measured at different times as a function
of cluster mass. In the absence of substructure, the amount of contraction
increases continuously with $M_{\rm cl}$, as expected from the mass
dependence of dynamical friction. In the clumpy halo, however, the curves
cross the initial half-light radius at a critical mass
$M_{\rm crit}\simeq10^5\msol$. Below $M_{\rm crit}$ the cluster population
expands, whereas above $M_{\rm crit}$ it contracts.

\begin{figure}
\begin{center}
\includegraphics[width=76mm]{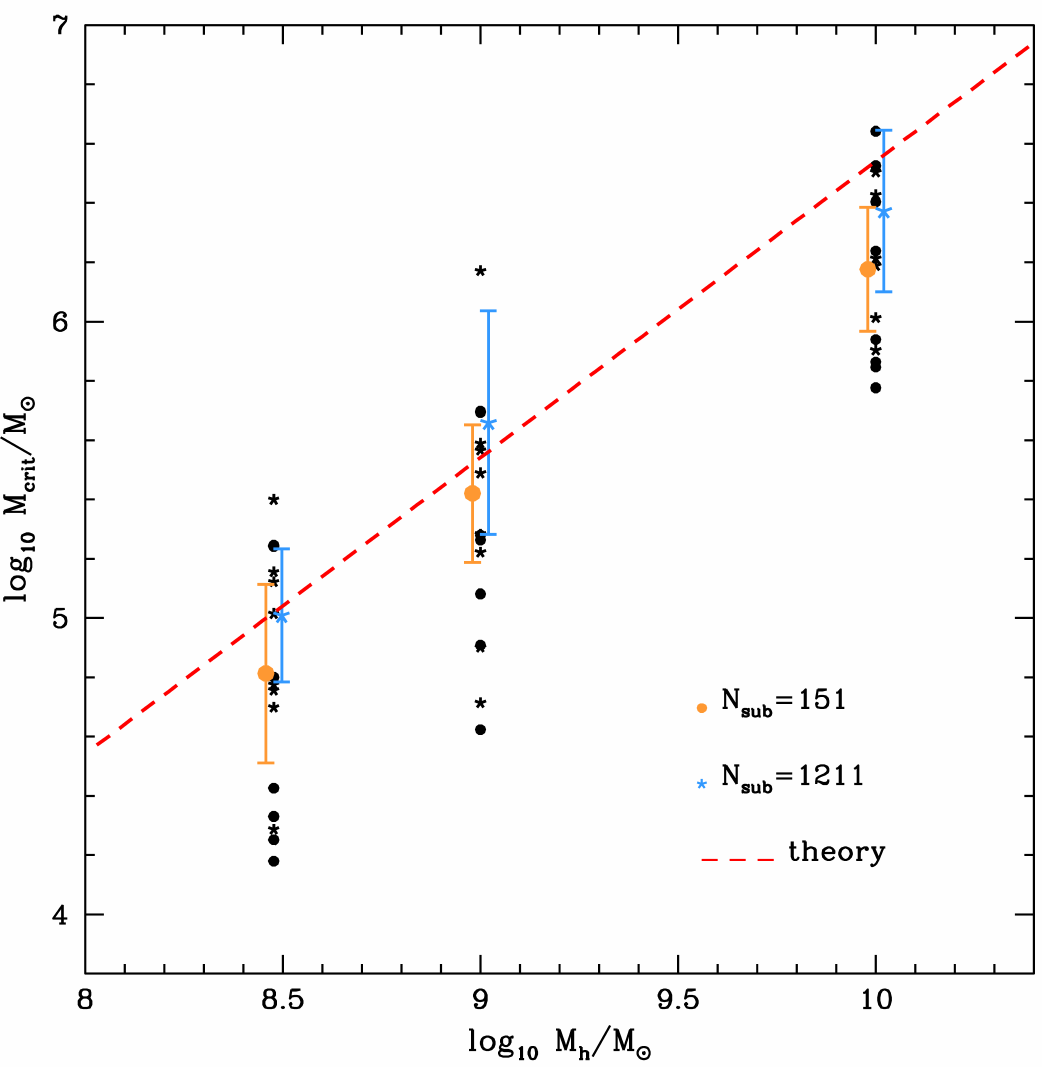}
\end{center}
\caption{Cluster critical mass as a function of the host halo mass. Individual subhalo realizations with a fixed mass-function normalization $B_0$ and low-mass cut-offs $\xi_1=10^{-4}$ ($N=151$) and $\xi_1=10^{-5}$ ($N=1211$) are shown with black dots and asterisks, respectively (see text). Averages between realizations are shown with coloured symbols, with error bars denoting the standard deviation around the mean. The dashed red line shows the analytical prediction from Equation~(\ref{eq:mcrit_CDM}) for $\sigma_{\rm cl}=\sigma$, adopting $\ln(\Lambda_{f,\rm h})=2.1$ and $\ln(\Lambda_{d,\rm sub})=8.2$, and the best-fitting $\ln(\Lambda_{f,\rm sub})=5.2$.}
\label{fig:meq}
\end{figure}

Fig.~\ref{fig:meq} shows $M_{\rm crit}$ measured from independent subhalo
realizations as a function of host halo mass. The numerical experiments
closely follow the linear scaling predicted by~(\ref{eq:mcrit})
over the range $3\times10^8\lesssim M_h/M_\odot\lesssim10^{10}$, although
they exhibit substantial scatter at fixed $M_h$. The origin of this scatter
can be traced to the upper end of the subhalo mass function. As shown by
Equation~(\ref{eq:mcrit_CDM}), stochastic
heating is very sensitive to the masses and orbits of the few largest perturbers in each realization.

As expected, decreasing the low-mass cut-off while keeping $B_0$ fixed
increases the total number of subhaloes but produces nearly identical
ensemble averages, confirming that the heating is controlled primarily by
the high-mass end of the CDM mass spectrum. We adopt
$\ln(\Lambda_{f,\rm h})=2.1$ for dynamical friction against the smooth halo
and the P19 value $\ln(\Lambda_{d,\rm sub})=8.2$ for stochastic heating,
and fit the simulations by varying $\ln(\Lambda_{f,\rm sub})$. The
simulations are accurately reproduced by Equation~(\ref{eq:mcrit}) with
$\ln(\Lambda_{f,\rm sub})\simeq5.2$, with a single value reproducing the
approximately linear relation over
$3\times10^8\lesssim M_h/M_\odot\lesssim10^{10}$. Interestingly, this
value is comparable to effective Coulomb logarithms measured in independent
numerical experiments. Errani et al. (2025) find
$\ln(\Lambda)\approx5$ using a direct-force integrator, whereas
Fellhauer \& Lin (2007) measure
$2\leq\ln(\Lambda)\leq5$ from collisionless simulations of a
point mass in an isothermal sphere.

\section{Discussion}\label{sec:discussion}

A particularly interesting application of our results concerns the globular cluster system of the Fornax dSph. We find $M_{\rm crit}\gtrsim10^5\msol$ (Fig.~\ref{fig:meq}), which is comparable to the masses of its clusters, suggesting that subhalo heating may significantly slow their orbital decay. 

An important simplification in our analysis is that the smooth and clumpy components of the halo are held fixed in time. In a cosmological setting, however, the stochastic force field and the corresponding critical mass will evolve. At early times, stellar feedback can generate baryonic fluctuations that heat dark matter and globular-cluster orbits, potentially transforming a central cusp into a core (Pontzen \& Governato 2012) and delaying cluster orbital decay. Interestingly, a constant-density core is favoured by chemo-dynamical modelling of Fornax (e.g. Walker \& Pe\~narrubia 2011; Pascale et al. 2018; Read et al. 2019). As star formation declines, subhaloes provide a long-lived source of secular heating. Since $M_{\rm crit}$ scales approximately with host halo mass (see Fig.~\ref{fig:meq}), halo growth increases the range of stellar-system masses for which stochastic heating can compete with dynamical friction. Conversely, Milky Way tides progressively lower $M_h$ and strip subhaloes, reducing force fluctuations (Pe\~narrubia \& Nadler 2026) and making dynamical friction comparatively more efficient. The heating--friction balance should therefore be regarded as a time-dependent property of the halo rather than a fixed threshold.

The low-halo mass regime may be particularly
important in the faintest dark-matter-dominated galaxies, where the number
of dark matter particles enclosed by stellar orbits and the characteristic
mass scales involved become small (Errani et al. 2025,2026). 

Furthermore, the mass dependence of the heating--friction balance
is bound to produce mass segregation among globular clusters in galactic DM haloes. Since the frictional energy loss increases with cluster mass, whereas stochastic heating is independent of $M_{\rm cl}$, massive clusters preferentially migrate inwards while lower-mass systems are heated outwards. Subhalo heating may therefore imprint a correlation between cluster mass and galactocentric distance, with the most massive clusters being preferentially found at smaller radii.

The mechanism considered here is not restricted to CDM. In fuzzy-dark-matter
models, for example, interference produces time-dependent density granules
whose gravitational fluctuations cause stochastic diffusion of stellar
orbits (El-Zant et al. 2020; Dutta Chowdhury et al. 2023). Competition with
dynamical friction may therefore produce an analogous critical mass. 
The detailed scaling of $M_{\rm crit}$, however, need not be the same as in
CDM because the amplitude, spatial scale and temporal correlations of the
fluctuating force are different.

Finally, the energy-balance argument only determines the direction of the
secular evolution. A natural extension is a Fokker--Planck description in
which gravitational encounters provide both drift and diffusion coefficients.
Stationary solutions with vanishing radial flux, $J=0$, may ultimately
predict the equilibrium spatial distribution and kinematics of globular
clusters as a function of their mass and the properties of the underlying
dark matter population. This will be explored in a separate contribution.

\section{Summary}\label{sec:sum}
In this work, we have explored the competition between dynamical friction
and stochastic heating in DM-dominated galaxies. The smooth DM component
systematically removes orbital energy from massive stellar systems through
dynamical friction, while subhaloes produce both a negative energy drift
and stochastic heating through gravitational encounters. Balancing these
contributions yields a critical mass, $M_{\rm crit}$, which separates two
qualitatively different regimes: systems with $M_{\rm cl}>M_{\rm crit}$
lose orbital energy and migrate inwards, whereas those with
$M_{\rm cl}<M_{\rm crit}$ are heated and expand within the host halo. At
$M_{\rm cl}\sim M_{\rm crit}$, heating and cooling approximately balance,
leading to no net expansion or contraction of the cluster population.

Our numerical experiments confirm this simple energy-balance argument. In dwarf galaxies, $M_{\rm crit}$ depends primarily on the abundance and mass spectrum
of dark subhaloes, as well as on the relative velocity dispersions of the
cluster and dark matter populations. For a CDM-like mass function with
$\alpha\simeq1.9$, the critical mass is controlled by the upper end of the
subhalo mass spectrum. Since these massive subhaloes are rare,
individual haloes exhibit substantial scatter around the mean prediction.

For dSph-sized CDM haloes we find $M_{\rm crit}\gtrsim10^5\msol$, comparable
to globular-cluster masses. Subhalo heating may therefore substantially delay their orbital
decay and contribute to the long-term survival of globular clusters in dwarf spheroidal galaxies. More generally, the competition between heating and friction predicts a mass-dependent spatial distribution of stellar systems in these galaxies.
Forthcoming wide-field surveys such as LSST, Euclid and Roman will greatly expand the census of globular clusters in dwarf galaxies, providing a potential new route to constrain the abundance and properties of otherwise invisible dark matter substructure in these galaxies.

\section*{Acknowledgements}
The authors wish to thank Luca Ciotti and Raphael Errani for their insightful comments.

\section*{Data availability}
No data were generated for this study.

{}

\end{document}